\documentclass[12pt]{iopart}
\usepackage{graphicx}
\usepackage{bm}
\usepackage{amssymb}
\usepackage{color}
\usepackage{latexsym}

\begin{document}

\title[Dynamic Properties of Motor Proteins]{Dynamic Properties of Motor Proteins with Two Subunits}
\author{Anatoly B. Kolomeisky and Hubert Phillips III}
\address{Department of Chemistry, Rice University, Houston, Texas 77251}

\begin{abstract}
The dynamics of motor protein molecules that have two subunits is investigated  using simple discrete stochastic models. Exact steady-state analytical expressions are obtained for  velocities and dispersions  for any number of  intermediate states and conformations between the corresponding binding states of proteins. These models enabled a detailed description and comparison of two possible mechanisms of the motion of motor proteins along the linear tracks: the hand-over-hand mechanism when the motion of subunits alternate, and the inchworm mechanism when one subunit is always trailing another one. It is shown that particles  in the hand-over-hand mechanism move faster and fluctuate more than the molecules in the inchworm mechanism. The effect of external forces on dynamic properties of motor proteins is discussed.  Finally, a method is proposed for distinguishing between these two mechanisms based on experimental observations. 
\end{abstract}

\ead{tolya@rice.edu}

\maketitle

\section{Introduction}

Motor proteins, also called molecular motors, are processive enzyme molecules that play a fundamental role in most biological processes, but especially in cellular transport, motility, cell division, and transcription \cite{lodish,bray,howard}. Motor proteins, such as kinesins, dyneins, myosins, polymerases, helicases, etc., function by stepping between equally spaced binding sites  along the rigid polar linear tracks (microtubules, actin filaments, DNA molecules), and the motion is powered by the hydrolysis of adenosine triphosphate (ATP) or related compounds. The mechanisms of the transformation of  chemical energy of hydrolysis  into the mechanical work in motor proteins are not yet fully understood \cite{howard}.  

Recent experimental advances have allowed for the  determination of structural and dynamic properties  of motor proteins  with a high-degree precision at a single-molecule level \cite{svoboda,kozielski,schnitzer,kojima,visscher,mehta,schnitzer1,nishiyama,asbury,yildiz,mallik,forkey,snyder}. Crystal structures suggest that many motor proteins  have two  domains elastically coupled together, each capable of hydrolyzing ATP and moving along the linear track \cite{howard,kozielski}. Two possible mechanisms of the coordinated motion of protein molecules with two motor heads have been proposed \cite{howard}. In a hand-over-hand mechanism, at each step only one  motor  head undergoes a sequence of mechanochemical transitions so that the motor subunits alternate between trailing and leading positions at the beginning of the cycle. In this mechanism the motor subunits are fully equivalent to each other.   In contrast, according to an inchworm mechanism, one motor domain is always ahead of another one during the cycle, i.e., the motor heads are not equivalent at all times.  Experiments on single-molecule fluorescently labeled myosins V, that step along the actin filaments, and on kinesins, that move along microtubules,  support the hand-over-hand mechanism for these motor proteins \cite{asbury,yildiz,forkey,snyder}. However, there are indications that dyneins probably utilize the inchworm mechanism \cite{oiwa}.

Successes in experimental studies  strongly stimulated many theoretical investigations of mechanisms and dynamics of molecular motors \cite{qian,julicher,KW,FK,FK1,KF00a,KF00b,kolom,FK2,mogilner,bustamante01,KF03,lan05}.  Most  theoretical work on motor proteins follows two main directions. One approach utilizes the concept of thermal ratchets \cite{julicher,bustamante01,lan05}. According to this idea, the motor protein is viewed as a Brownian particle that moves in two different periodic but asymmetric potentials, switching stochastically between them. This method  takes into account the internal structure and interactions between different  domains in protein molecules, however, the results are mainly numerical and  depend on the specific potentials used in calculations. An alternative approach is based on multi-state discrete stochastic (chemical kinetic) models \cite{qian,KW,FK,FK1,KF00a,KF00b,kolom,FK2,KF03}.  In this method the molecular motors are associated with  particles that move along one-dimensional periodic lattices with different forward and backward rates. The lattices corresponds to  biochemical pathways for the motor proteins, while the sites in the period describe the biochemical cycle when the protein molecule travels between two consecutive binding sites. Using this mapping of the motion of a random walker and applying the method of Derrida \cite{derrida83}, {\it exact analytical} expressions for the mean velocities and dispersions are derived for {\it any} number of intermediate states (i.e., for  the period of any size) and for different complexity of  biochemical pathways \cite{FK,FK1,KF00a,KF00b,kolom}. It was demonstrated that this approach allowed for the successful analysis of  the full dynamics of single kinesin and myosin V molecules \cite{FK2,KF03}.  However, the weakness of this method is the fact that the internal structure of motor proteins, namely, the motion and interactions of different subunits, is not taken into account.

Determining  how the different motor heads move relative to each other  is critical for the overall understanding of motor protein's dynamics and functions. Current experimental methods with single-molecule fluorescent labels, that  distinguish between the different types of molecular motion,  require a detailed knowledge of the protein structure which is not always available. In addition, the labeled proteins may change their biochemical properties in comparison with the original species. However, it would be more advantageous to use simpler less-invasive experimental methods along with better theoretical models to study the specific mechanisms of molecular motors.  In this article we investigate the dynamics of motor proteins by developing a set of simple multi-state discrete stochastic models. In our approach the motor proteins consist of two interacting particles that correspond to different motor subunits in real enzymes. Explicit formulas for the velocities and dispersions are obtained for two different mechanisms of motion. We  suggest a method  to distinguish between two possible mechanisms by analyzing  time trajectories of single motor proteins obtained in optical-trap experiments \cite{schnitzer,kojima,visscher,mehta,schnitzer1,nishiyama} in combination with the bulk biochemical kinetic data.

\section{Theoretical Approach}

\subsection{General Model}

Consider a motor protein molecule with two subunits that travels along the filament track. We model this system as two identical interacting particles moving on a periodic one-dimensional lattice, as shown in Fig. 1. There are $N$ intermediate discrete states on a biochemical pathway between two consecutive binding sites. In the simplest approximation, we assume that the particles interact through hard-core exclusion, i.e., they cannot occupy the same site. Also, the particles cannot run away from each other. If $x_{1}$ and $x_{2}$ are the positions of the motor particles then
\begin{equation} \label{condition}
|x_{1}-x_{2}| \le m, 
\end{equation}
where  $m$ is an integer that specifies how far apart two motor domains can be found in the protein molecule. In the lattice the sites $x=\pm N l$ ($l=0, 1, \cdots$) correspond to binding sites of the molecular motor. The distance between two consecutive binding sites is $d$, which is equal to 8.2 nm for kinesins and dyneins moving on microtubules and 36 nm for myosins V and VI traveling along the actin filaments \cite{lodish,bray,howard}.

\begin{figure}[tbp]
\centering
\includegraphics[clip=true]{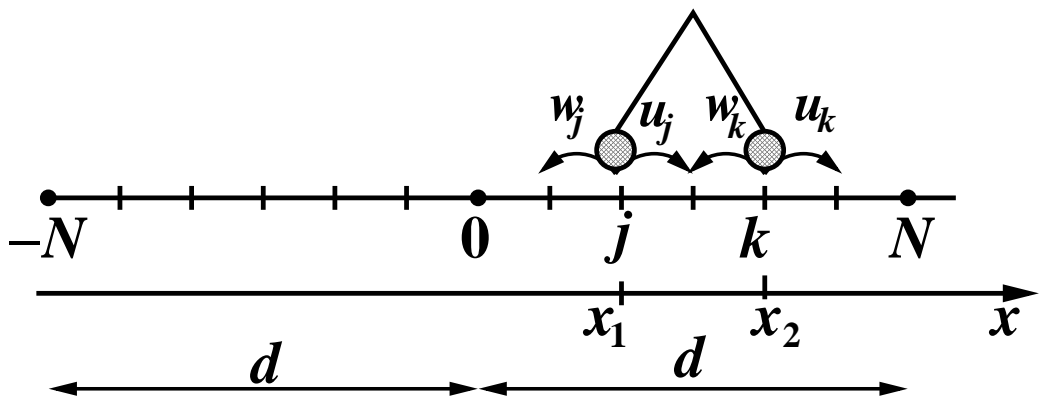}
\caption{General schematic view of periodic stochastic models of motor proteins consisting of two  subunits. Two parts of the molecular motor cannot occupy the same site and cannot be more than $m$ sites apart. The motor domain at site $j$ can make a forward or backward steps with the rate $u_{j}$ or $w_{j}$, correspondingly, if these transitions are allowed by another motor domain.}
\label{fig1}
\end{figure}

The particle at site $j$ moves forward (backward) with the rate $u_{j}$ ($w_{j}$) if the site $j+1$ ($j-1$) is available and the move does not violate the condition (\ref{condition}): see Fig. 1. Because of periodicity the transition rates are related, $u_{j \pm N l}=u_{j}$ and  $w_{j \pm N l}=w_{j}$ for $l=0,1 \cdots$ and $0 \le j \le N-1$. The dynamic properties of motor proteins  are specified by the drift velocity
\begin{equation}\label{eq.V}
V=V(\{ u_{j},w_{j} \}) = \lim_{ t \rightarrow \infty} \frac{d}{d t} \langle x(t) \rangle,
\end{equation}
and dispersion (or diffusion constant)
\begin{equation}\label{eq.D}
D=D(\{ u_{j},w_{j} \}) = \lim_{ t \rightarrow \infty} \frac{1}{2} \frac{d}{d t} \left[ \langle x^{2}(t) \rangle - \langle x(t) \rangle^{2} \right],
\end{equation}
where $x(t)$ is the position of the center of mass of the protein molecule at time $t$.   It is convenient to express the degree of fluctuations of the molecular motor in terms of a dimensionless function called randomness \cite{svoboda}
\begin{equation}
r=\frac{2 D}{d V}.
\end{equation}
This function  sets bounds on the number of rate-limiting biochemical transitions and thus yields an important information about the mechanism of a motor protein's processivity \cite{visscher,KF00a,FK2}.

The motor proteins in experiments and in the cellular environment frequently work against external loads \cite{lodish,bray,howard}. External forces modify the transition rates in the following way \cite{FK,FK1,KF00a,KF00b,kolom}
\begin{equation}\label{eq.load}
u_{j}(F)=u_{j}(0) \exp \left( -\frac{ \theta_{j}^{+} F d}{ k_{B}T} \right), \quad w_{j}(F)=w_{j}(0) \exp \left( +\frac{ \theta_{j}^{-} F d}{ k_{B}T} \right),
\end{equation}
where $\sum_{j=0}^{N-1}( \theta_{j}^{+} + \theta_{j}^{-})=1$, and  $\theta_{j}^{\pm}$ are load-distribution factors that specify how the external load changes the energy  activation
 barriers for the biochemical transitions from the state $j$.

The dynamic properties of the motor proteins with two domains depend on the specific mechanism of the motion. Below we consider in detail the hand-over-hand and the inchworm mechanisms.

\subsection{Hand-Over-Hand Mechanism}

In this mechanism, the trailing subunit makes $N$ intermediate steps and becomes the leading particle, as shown in Fig. 2. Then the next particle makes $N$ transitions. During the cycle each head advances the distance $2d$ so the center of mass of the protein molecule moves only the distance $d$. This mechanism then can be viewed as a motion of two particles on periodic parallel  one-dimensional lattices, where the distance between neighboring binding sites is $2d$. Because the particles are identical, this picture is easily mapped into the motion of the single particle (center of mass) on the original one-dimensional lattice  for which the dynamics is well understood \cite{derrida83,FK}.

\begin{figure}[tbp]
\centering
\includegraphics[clip=true]{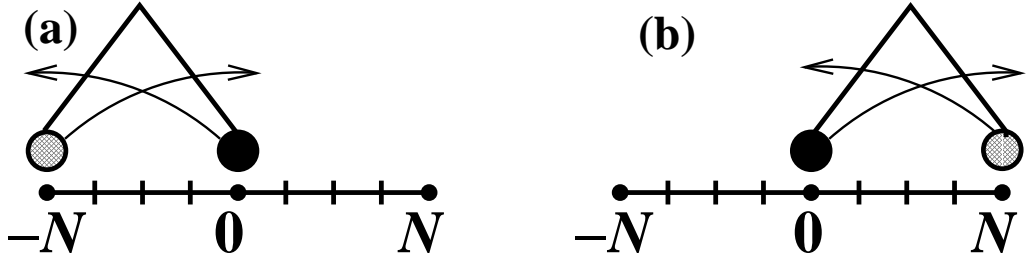}
\caption{The general picture of the hand-over-hand mechanism for motor proteins with two heads. (a) The black particle is a leading motor domain and it cannot move forward. The grey subunit is  passing over  the black subunit. (b) The black particle is now the trailing subunit and it is available for the forward motion. Arrows indicate the allowed transitions.}
\label{fig2}
\end{figure}

At large times the exact expression for the drift velocity is given by \cite{derrida83,FK}
\begin{equation}
V_{hoh}=d \left( 1-\prod_{j=0}^{N-1} \frac{w_{j}}{u_{j}} \right)/R_{N},
\end{equation}
and
\begin{equation}
R_{N}=\sum_{j=0}^{N-1} r_{j}, \quad r_{j}=u_{j}^{-1} \left[ 1+ \sum_{k=1}^{N-1} \prod_{j+1}^{j+k} \frac{w_{i}}{u_{i}} \right].
\end{equation}
The corresponding expression for dispersion can be written as  \cite{derrida83,FK}
\begin{equation}
D_{hoh}=\frac{d}{N} \left( \frac{d U_{N}+V S_{N}}{R_{N}^{2}} -\frac{(N+2)V}{2} \right),
\end{equation}
where the auxiliary functions are
\begin{equation}
S_{N}=\sum_{j=0}^{N-1} s_{j} \sum_{k=0}^{N-1} (k+1) r_{k+j+1}, \quad U_{N}=\sum_{j=0}^{N-1} u_{j} r_{j} s_{j}, 
\end{equation}
and
\begin{equation}
s_{j}=u_{j}^{-1} \left[ 1+ \sum_{k=1}^{N-1} \prod_{j-1}^{j-k} \frac{w_{i+1}}{u_{i}} \right].
\end{equation}
It can also be demonstrated  that in  this mechanism  $r_{hoh} > 1/N$ for any set of transitions rates $\{u_{j},w_{j} \}$ \cite{visscher,KF00a,FK2}.

\subsection{Inchworm Mechanism}

Now consider the inchworm mechanism of the motion of the motor proteins with two subunits. In this mechanism one particle is always leading the other one, as shown in Fig. 3. In the simplest approximation, we discuss only the case  $m=2$ in Eq. (\ref{condition}). i.e., the two motor particles can only be found on two nearest-neighbor sites or next-nearest-neighbor sites: see Fig. 3.  

\begin{figure}[tbp]
\centering
\includegraphics[clip=true]{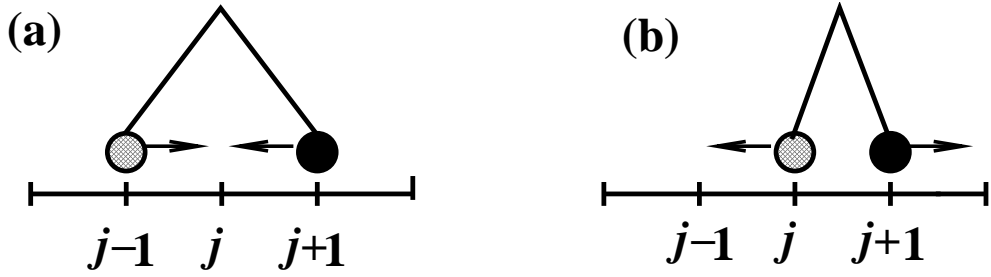}
\caption{Two possible configurations for the inchworm mechanism of the motion with $m=2$. The black particle is always leading. The allowed transitions are indicated by arrows.}
\label{fig3}
\end{figure}

To determine the dynamic properties of motor proteins in this model we develop a method that generalizes the original Derrida's approach \cite{derrida83,KF00a,KF00b}. The first step is to introduce $P_{j,k}(l,t)$ which is the probability to find the trailing subunit of the motor protein molecule at state $j$  and the leading subunit at state $k$  ($k=j+1$ or $j+2$) at site $l$  at time $t$ (see Fig. 3). The time evolution of this probability is governed by Master equations:
\begin{eqnarray} \label{master.eq}
\frac{dP_{j,j+1}(l,t)}{dt} & = & u_{j-1}P_{j-1,j+1}(l,t)+w_{j+2}P_{j,j+2}(l,t) \nonumber \\ 
                           &   & -\left(u_{j+1}+w_{j}\right)P_{j,j+1}(l,t)     \nonumber \\ 
\frac{dP_{j,j+2}(l,t)}{dt} & = & u_{j+1}P_{j,j+1}(l,t)+w_{j+1}P_{j+1,j+2}(l,t) \nonumber \\ 
                           &   & -\left(u_{j}+w_{j+2}\right)P_{j,j+2}(l,t). 
\end{eqnarray}
Because of the conservation of probability, we require
\begin{equation}
\sum_{l=-\infty}^{+\infty} \sum_{j=0}^{N-1}\left[P_{j,j+1}(t)+P_{j,j+2}(l,t)\right]=1 \quad $   for all $t.
\end{equation}
Next following  Derrida's method \cite{derrida83}, we define auxiliary functions
\begin{equation}
B_{j,j+1}(t)\equiv\sum_{l=-\infty}^{+\infty}P_{j,j+1}(l,t), \quad B_{j,j+2}(t)\equiv\sum_{l=-\infty}^{+\infty}P_{j,j+2}(l,t),
\end{equation}
and
\begin{eqnarray}
C_{j,j+1}(t)\equiv\sum_{l=-\infty}^{+\infty}\left(j+Nl\right)P_{j,j+1}(l,t), \nonumber \\
C_{j,j+2}(t)\equiv\sum_{l=-\infty}^{+\infty}\left(j+Nl\right)P_{j,j+2}(l,t).  
\end{eqnarray}
Using the Master equations (\ref{master.eq}) we derive
\begin{eqnarray}\label{B.eq}
\frac{d}{dt}B_{j,j+1}(t)=u_{j-1}B_{j-1,j+1}(t)+w_{j+2}B_{j,j+2}(t)-\left(u_{j+1}+w_{j}\right)B_{j,j+1}(t), \nonumber \\
\frac{d}{dt}B_{j,j+2}(t)=u_{j+1}B_{j,j+1}(t)+w_{j+1}B_{j+1,j+2}(t)-\left(u_j+w_{j+2}\right)B_{j,j+2}(t); 
\end{eqnarray}
and
\begin{eqnarray}\label{C.eq}
\frac{d}{dt}C_{j,j+1}(t) & = & u_{j-1}C_{j-1,j+1}(t)+w_{j+2}C_{j,j+2}(t)  -\left(u_{j+1}+w_{j}\right)C_{j,j+1}(t) \nonumber \\
                         &   & + u_{j-1}B_{j-1,j+1}(t) \nonumber \\
\frac{d}{dt}C_{j,j+2}(t)& = & u_{j+1}C_{j,j+1}(t)+w_{j+1}C_{j+1,j+2}(t)  -\left(u_{j}+w_{j+2}\right)C_{j,j+2}(t) \nonumber \\
                        &   &+ w_{j+1}B_{j+1,j+2}(t).
\end{eqnarray}

In the limit of $t \rightarrow \infty$, again following Derrida's suggestions \cite{derrida83}, we introduce the ansatz  
\begin{eqnarray}\label{ansatz}
B_{j,k}(t)\rightarrow b_{j,k}, \quad  C_{j,k}(t)-a_{j,k}t\rightarrow T_{j,k}.
\end{eqnarray}
Note that the parameters $b_{j,k}$, $a_{j,k}$ and $T_{j,k}$  are periodic, i.e., $b_{j,k}=b_{j+N,k+N}$, $a_{j,k}=a_{j+N,k+N}$, and $T_{j,k}=T_{j+N,k+N}$.  Now  define two new functions, 
\begin{equation}\label{eq.f}
f_{j-1}^{(1)}\equiv w_{j}b_{j,j+1}-u_{j-1}b_{j-1,j+1}, \quad  f_{j+1}^{(2)}\equiv w_{j+2}b_{j,j+2}-u_{j+1}b_{j,j+1}.
\end{equation}
At steady  state  $\frac{dB_{j,k}(t)}{dt}=0$, and Eqs. (\ref{B.eq}) transform into
\begin{eqnarray}\label{eq.b}
0 = u_{j-1}b_{j-1,j+1} + w_{j+2}b_{j,j+2} - \left( u_{j+1} + w_{j} \right) b_{j,j+1}, \nonumber \\
0 = u_{j+1}b_{j,j+1} + w_{j+1}b_{j+1,j+2} - \left( u_{j} + w_{j+2} \right) b_{j,j+2}.
\end{eqnarray}
Substituting (\ref{eq.f}) into these equations,  we obtain 
\begin{eqnarray}
f_{j}^{(1)}=w_{j+1}b_{j+1,j+2}-u_{j}b_{j,j+2}=f_{0}, \nonumber \\ 
f_{j+1}^{(2)}=w_{j+2}b_{j,j+2}-u_{j+1}b_{j,j+1}=f_{0},
\end{eqnarray}
where $f_{0}$ is a constant. Then it can be shown that $f_{j}^{(1)}=f_{j-1}^{(1)}=f_{0}=f_{j}^{(2)}=f_{j+1}^{(2)}$.  This leads to the following expression for $b_{j,k}$ 
\begin{eqnarray}
b_{j,j+1}=\frac{-f_{0}}{u_{j+1}}+\frac{w_{j+2}}{u_{j+1}}b_{j,j+2} 
= \frac{-f_{0}}{u_{j+1}}\left[1 + \frac{w_{j+2}}{u_{j}}\right] + \frac{w_{j+1}w_{j+2}}{u_{j}u_{j+1}}b_{j+1,j+2}, \nonumber \\
b_{j,j+2}=\frac{-f_{0}}{u_{j}}+\frac{w_{j+1}}{u_{j}}b_{j+1,j+2}
= \frac{-f_{0}}{u_{j}}\left[1 + \frac{w_{j+1}}{u_{j+2}}\right] + \frac{w_{j+1}w_{j+3}}{u_{j}u_{j+2}}b_{j+1,j+3}. 
\end{eqnarray}
Solving these equations by iteration, and using the periodicity and  the normalization condition, 
\begin{equation}\label{eq.norm}
\sum_{j=0}^{N-1}\left(b_{j,j+1}+b_{j,j+2}\right)=1,
\end{equation}
 we finally derive
\begin{equation}\label{eq.R}
   b_{j,k} = \frac{r_{j,k}}{R_{N}}, \quad R_N = \sum_{j=0}^{N-1} \left[ r_{j,j+1} + r_{j,j+2} \right],
\end{equation}
where
\begin{eqnarray}\label{eq.r}
r_{j,j+1} = \frac{1}{u_{j+1}} \left\{ 1 + \sum_{k=1}^{N-1} \prod_{i=j}^{j+k-1} \left( \frac{w_{i+1}w_{i+2}}{u_{i}u_{i+2}} \right) + \frac{w_{j+2}}{u_{j}} \left[ 1 + \sum_{k=1}^{N-1} \prod_{i=j}^{j+k-1} \left( \frac{ w_{i+1} w_{i+3} }{ u_{i+1} u_{i+2} } \right) \right] \right\}, \nonumber \\ 
r_{j,j+2} = \frac{1}{u_{j}} \left\{ 1 + \sum_{k=1}^{N-1} \prod_{i=j}^{j+k-1} \left( \frac{w_{i+1}w_{i+3}}{u_{i+1}u_{i+2}} \right) + \frac{w_{j+1}}{u_{j+2}} \left[ 1 + \sum_{k=1}^{N-1} \prod_{i=j}^{j+k-1} \left( \frac{w_{i+2}w_{i+3}}{u_{i+1}u_{i+3}} \right) \right] \right\}. 
\end{eqnarray}

To determine the coefficients $a_{j,k}$ and $T_{j,k}$, the ansatz (\ref{ansatz}) is substituted into Eqs. (\ref{C.eq}) in the limit of large times. This yields the following equations
\begin{eqnarray}\label{eq.a}
0 = u_{j-1}a_{j-1,j+1} + w_{j+2}a_{j,j+2} - \left( u_{j+1} + w_{j} \right) a_{j,j+1}, \nonumber \\
0 = u_{j+1}a_{j,j+1} + w_{j+1}a_{j+1,j+2} - \left( u_{j} + w_{j+2} \right) a_{j,j+2}.
\end{eqnarray}
Also the parameters  $T_{j,k}$ must satisfy 
\begin{eqnarray}\label{eq.T}
a_{j,j+1} = u_{j-1}T_{j-1,j+1} + w_{j+2}T_{j,j+2} - \left( u_{j+1} + w_{j} \right) T_{j,j+1} - u_{j-1}b_{j-1,j+1}, \nonumber \\
a_{j,j+2} = u_{j+1}T_{j,j+1} + w_{j+1}T_{j+1,j+2} - \left( u_{j} + w_{j+2} \right) T_{j,j+2} - w_{j+1}b_{j+1,j+2}.
\end{eqnarray}
Comparing Eqs. (\ref{eq.a}) with Eqs. (\ref{eq.b}) we conclude that $a_{j,k}=A b_{j,k}$. The coefficient $A$ can be found using  the normalization condition (\ref{eq.norm}) and it is equal to
\begin{equation}\label{eq.A}
 A=\sum_{j=0}^{N-1}\frac{u_{j}r_{j,j+2}-w_{j}r_{j,j+1}}{R_N} = N \frac{1 - \left( \prod_{j=0}^{N-1} \frac{w_{j}}{u_{j}} \right)^2}{R_N}.
\end{equation}
To calculate the coefficients $T _{j,k}$ we introduce another set of auxiliary functions
\begin{equation}
   y^{(1)}_{j+1} \equiv w_{j+2}T_{j,j+2}-u_{j+1}T_{j,j+1}, \quad y^{(2)}_{j-1} \equiv w_{j}T_{j,j+1}-u_{j-1}T_{j-1,j+1}.
\end{equation}
Then Eqs. (\ref{eq.T}) can be rewritten in the following form
\begin{eqnarray}
a_{j,j+1} = y^{(1)}_{j+1} - y^{(2)}_{j-1} + u_{j-1}b_{j-1,j+1}, \nonumber \\
a_{j,j+2} = y^{(2)}_{j} - y^{(1)}_{j+1} - w_{j+1}b_{j+1,j+2}.
\end{eqnarray}
As shown in \cite{KF00a,KF00b},  these  equations are solved to yield  the functions $y^{(1)}_{j}$ and $y^{(2)}_{j}$ 
\begin{eqnarray}
y^{(1)}_{j} = -a_{j-1,j+1} + \frac{A}{N} \left( 1+ \sum_{i=0}^{N-1} \left( i+1 \right) \left[ b_{j+i,j+i+1} + b_{j+i,j+i+2} \right] \right) + C_{1}, \nonumber \\
y^{(2)}_{j} = u_{j}b_{j,j+2} + \frac{A}{N}  \left [ \sum_{i=0}^{N-1} \left( i+1 \right)  b_{j+i+1,j+i+2} + b_{j+i+1,j+i+3} \right] + C_{2},
\end{eqnarray}
where the coefficients $C_{1}$ and $C_{2}$ are arbitrary constants that cancel in the final expression for  dispersion. Then the final expressions for $T_{j,k}$ are found to be  \cite{KF00a,KF00b}
\begin{eqnarray}
T_{j,j+1} \left[ 1- \left( \prod_{j=0}^{N-1} \frac{w_j}{u_j} \right)^2 \right] = \frac{-1}{u_{j+1}} \left[ y^{(1)}_{j+1} + \sum_{k=1}^{N-1} y^{(1)}_{j+k+1} \prod_{i=j}^{j+k-1} \left( \frac{w_{i+1}w_{i+2}}{u_{i}u_{i+2}} \right) \right. \nonumber \\
 \left.  + \frac{w_{j+2}}{u_{j}} \left( y^{(2)}_{j} + \sum_{k=1}^{N-1} y^{(2)}_{j+k} \prod_{i=j}^{j+k-1} \left( \frac{w_{i+1}w_{i+3}}{u_{i+1}u_{i+2}} \right) \right) \right]; \nonumber \\
T_{j,j+2} \left[ 1- \left( \prod_{j=0}^{N-1} \frac{w_j}{u_j} \right)^2 \right] = \frac{-1}{u_{j}} \left[ y^{(2)}_{j} + \sum_{k=1}^{N-1} y^{(2)}_{j+k} \prod_{i=j}^{j+k-1} \left( \frac{w_{i+1}w_{i+3}}{u_{i+1}u_{i+2}} \right) \right. \nonumber \\
 \left.  + \frac{w_{j+1}}{u_{j+2}} \left( y^{(1)}_{j+2} + \sum_{k=1}^{N-1} y^{(1)}_{j+k+2} \prod_{i=j}^{j+k-1} \left( \frac{w_{i+2}w_{i+3}}{u_{i+1}u_{i+3}} \right) \right) \right].
\end{eqnarray}

For simplicity the trailing subunit is chosen as a marker for derivation  the explicit expressions for the drift velocity and dispersion. It can be shown that the same results are obtained if the center of mass is used. The position of this particle at any time is given by
\begin{eqnarray}
   \left<x(t)\right> =  \frac{d}{N}\sum_{l=-\infty }^{\infty }\sum_{j=0}^{N-1}\left( j + Nl \right) \left[ P_{j,j+1}(l,t) + P_{j,j+2}(l,t) \right]  \nonumber \\
 = \frac{d}{N}\sum_{j=0}^{N-1}\left[ C_{j,j+1}(t) + C_{j,j+2}(t) \right].
\end{eqnarray}
Utilizing  this result along with the  Master equations (\ref{master.eq})  we get  
\begin{eqnarray}
 \frac{d\left<x(t)\right>}{dt}= \frac{d}{N}\sum_{l=-\infty}^{\infty}\sum_{j=0}^{N-1}(j+Nl) \left[ u_{j-1}P_{j-1,j+1}(l,t) \right. \nonumber \\
 \left.  -w_{j}P_{j,j+1}(l,t)+w_{j+1}P_{j+1,j+2}(l,t)-u_{j}P_{j,j+2}(l,t) \right ].
\end{eqnarray}
Then the average drift velocity (\ref{eq.V}) has a simple form $V=\frac{d}{N}A$ where the function $A$ is given by Eq. (\ref{eq.A}). The final expression for the velocity is 
\begin{equation}\label{eq.velocity}
 V=d\frac{\left[1-\left(\prod_{j=0}^{N-1}\frac{w_{j}}{u_{j}}\right)^2\right]}{R_{N}}.
\end{equation}

A similar approach can be used to determine the dispersion, which can be done  with the help of the following equation, 
\begin{equation}
\left< x^2(t) \right> = \frac{d^2}{N^2} \sum_{l=-\infty}^{+\infty} \sum_{j=0}^{N-1} \left( j+Nl \right)^2 \left[ P_{j,j+1}(l,t) + P_{j,j+2}(l,t) \right].
\end{equation}
The time evolution of this quantity, again applying the Master equations (\ref{master.eq}), is given by
\begin{equation}
\frac{d \left< x^2(t) \right>}{dt} = 2 \left( \frac{d}{N} \right)^2 \sum_{j=0}^{N-1} \left[ u_{j}C_{j,j+2}(t) - w_{j}C_{j,j+1} + \frac{1}{2} \left( u_{j}B_{j,j+2} + w_{j}B_{j,j+1} \right) \right]
\end{equation}
Then after some algebra the expression for dispersion is written as
\begin{eqnarray}
D = \left( \frac{d}{N} \right)^2 \sum_{j=0}^{N-1} \left[  u_{j}T_{j,j+2} - w_{j}T_{j,j+1}  + \frac{1}{2} \left( u_{j}b_{j,j+2} + w_{j}b_{j,j+1} \right) \right. \nonumber \\
 \left. - A ( T_{j,j+1} + T_{j,j+2}) \right].
\end{eqnarray}
And using the definition (\ref{eq.D}) we obtain the final formula: 
\begin{equation}\label{eq.dispersion}
D = \frac{d}{N} \left( \frac{dU_{N}+VS_{N}}{R_{N}^{2}}-\frac{N+2}{2}V \right),
\end{equation}
where
\begin{eqnarray}\label{eq.disp1}
U_{N} = \sum_{j=0}^{N-1} \left[ s_{2}(j)u_{j}r_{j,j+2} + s_{1}(j) \frac{A}{N} \left( R_{N} - Nr_{j-1,j+1} \right) \right], \nonumber \\
S_{N} = \sum_{j=0}^{N-1} \left( s_{1}(j) + s_{2}(j-1) \right) \sum_{i=0}^{N-1} \left(i+1\right) \left[ r_{j+i,j+i+1} + r_{j+i,j+i+2} \right], \nonumber \\
s_{1}(j) = \frac{1}{u_{j}} \left[ 1 + \sum_{k=1}^{N-1} \prod_{i=j}^{j-k+1} \frac{w_{i-1}w_{i}}{u_{i-2}u_{i-1}} + \frac{w_{j-1}}{u_{j-2}} \left( 1 + \sum_{k=1}^{N-1} \prod_{i=j}^{j-k+1}  \frac{w_{i-2}w_{i}}{u_{i-3}u_{i-1}}  \right) \right], \nonumber \\
s_{2}(j) = \frac{1}{u_{j}} \left[ 1 + \sum_{k=1}^{N-1} \prod_{i=j}^{j-k+1} \frac{w_{i}w_{i+2}}{u_{i-1}u_{i+1}} + \frac{w_{j+2}}{u_{j+1}} \left( 1 + \sum_{k=1}^{N-1} \prod_{i=j}^{j-k+1}  \frac{w_{i}w_{i+1}}{u_{i-1}u_{i}}  \right) \right].
\end{eqnarray}

In this mechanism  the bounds on the randomness parameter $r$ can be estimated by calculating the dynamic properties for the simple limiting case when $u_{j}=u$ and $w_{j}=0$ for all $j$. Then  Eqs. (\ref{eq.R}), (\ref{eq.r}) and  (\ref{eq.velocity}) yield
\begin{equation}
R_{N}=\frac{2N}{u}, \quad V_{inch}=d \frac{u}{2N},
\end{equation}
while for dispersion we get from Eqs. (\ref{eq.dispersion}) and (\ref{eq.disp1})
\begin{equation}
S_{N}=\frac{2 N^{2}(N+1)}{u^{2}}, \quad U_{N}=\frac{3N}{2u}, \quad D_{inch}=\left( \frac{d}{N} \right)^{2} \frac{u}{8}.
\end{equation}
This analysis leads to $r=1/2N$ which is the smallest possible value for this parameter. For any other set of transition rates $\{u_{j},w_{j} \}$ the velocity is always smaller and the dispersion is larger, giving the general inequality for the inchworm mechanism
\begin{equation}
r_{inch} \ge \frac{1}{2N}.
\end{equation}

Although we considered here only the case of $m=2$, the method can be extended to include the inchworm  models where the particles can be found more than 2 sites apart.

\subsection{Comparison of Two Mechanisms}

The existence of exact analytical expressions for the dynamic properties of motor proteins with two heads in the hand-over-hand and the inchworm mechanisms allows us to analyze and compare these mechanisms very efficiently.

Consider first the simplest $N=1$ models. Then the average velocity and dispersion for the hand-over-hand mechanism is given by
\begin{equation}
V_{hoh}=d(u-w), \quad D_{hoh}=d^{2} (u+w)/2,
\end{equation}
The corresponding expressions for the inchworm mechanism can be obtained from Eqs. (\ref{eq.R}), (\ref{eq.r}), (\ref{eq.velocity}), (\ref{eq.dispersion}) and (\ref{eq.disp1})
\begin{equation}
V_{inch}=d(u-w)/2, \quad D_{inch}=d^{2} (u+w)/8.
\end{equation}
Thus the mean velocity in the inchworm model is only half of the velocity in the hand-over-hand mechanism, while the inchworm dispersion is only a quarter of the hand-over-hand value.

More interesting  case is $N=2$ models where the average velocity and dispersion for the hand-over-hand mechanism are \cite{derrida83,KW,FK}
\begin{equation}
V_{hoh}=d(u_{0}u_{1}-w_{0}w_{1})/\sigma, \ D_{hoh}=\frac{1}{2} d^{2} \left[ (u_{0}u_{1}+w_{0}w_{1}) -2 (V_{hoh}/d)^{2} \right] /\sigma, 
\end{equation}
where $\sigma=u_{0}+u_{1}+w_{0}+w_{1}$. For the inchworm mechanism the expression for the mean velocity can be written as
\begin{equation}
V_{inch}=d \ \frac{(u_{0}u_{1})^{2}-(w_{0}w_{1})^{2}} {2 \sigma (u_{0}u_{1}+w_{0}w_{1}) + (u_{0}w_{0}-u_{1}w_{1})(u_{0}+w_{0}-u_{1}-w_{1})},
\end{equation}
while the explicit formula for dispersion is very bulky and it will not be presented here. Instead, we analyze the dependence of the dynamics properties of motor proteins on external forces using Eqs. (\ref{eq.load}). 

Force-velocity curves for different mechanisms are presented in Fig. 4. It can be seen that the velocity for the inchworm mechanism is  always smaller then the corresponding curve for the hand-over-hand mechanism, although the stall forces are the same. This can be explained by recalling that the stall force is a thermodynamic parameter for the sequential chemical kinetic models \cite{FK}. It is equal to the free energy difference between two consecutive binding sites divided over the step size $d$. Both the free energy difference and the step size  are the same for the hand-over-hand and the inchworm mechanism, and this leads to the same value of the stall force.  

\begin{figure}[tbp]
\centering
\includegraphics[clip=true]{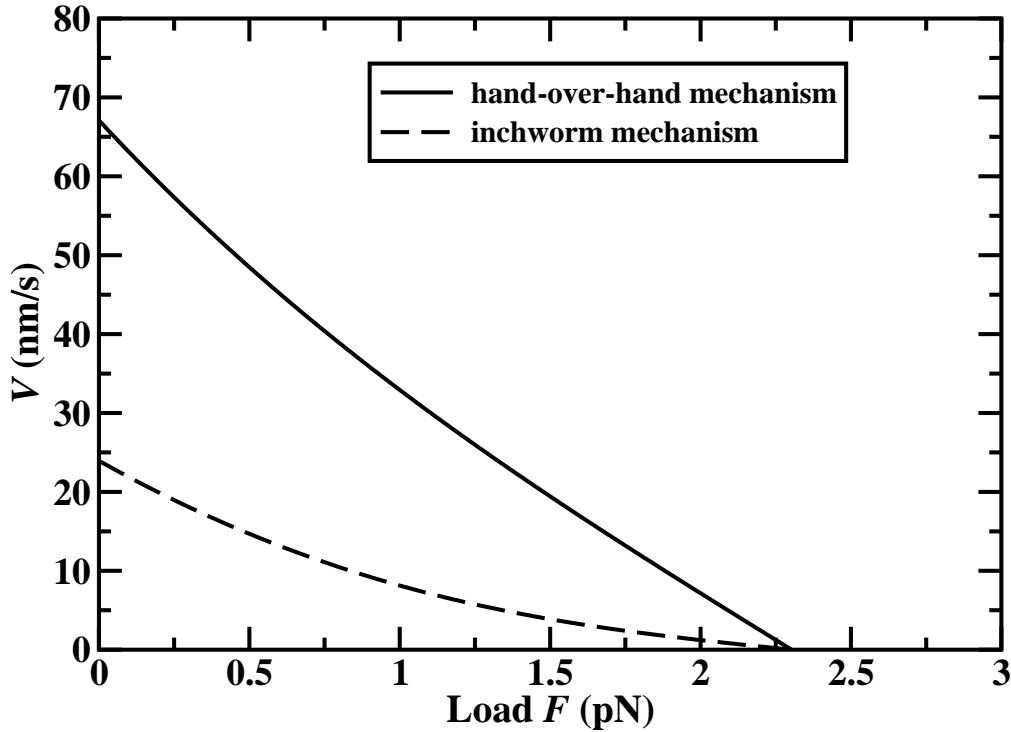}
\caption{Force-velocity curves for $N=2$ model with $u_{0}=10$ s$^{-1}$, $u_{1}=100$ s$^{-1}$, $w_{0}=1$ s$^{-1}$, $w_{1}=10$ s$^{-1}$, $\theta_{0}^{+}=\theta_{0}^{-}=\theta_{1}^{+}=\theta_{1}^{-}=0.5$ and $d=8.2$ nm. The solid line describes the hand-over-hand mechanism, while the dashed line corresponds to the inchworm mechanism.}
\label{fig4}
\end{figure}

The properties of dispersions for two mechanisms at different external loads, as shown in Fig. 5, are similar to the mean velocities. The particles that move via the inchworm mechanism fluctuate much less than the motor proteins utilizing the hand-over-hand method. This behavior is expected since one of the motor subunits lowers the stochastic fluctuations of another motor head in the inchworm mechanism.

\begin{figure}[tbp]
\centering
\includegraphics[clip=true]{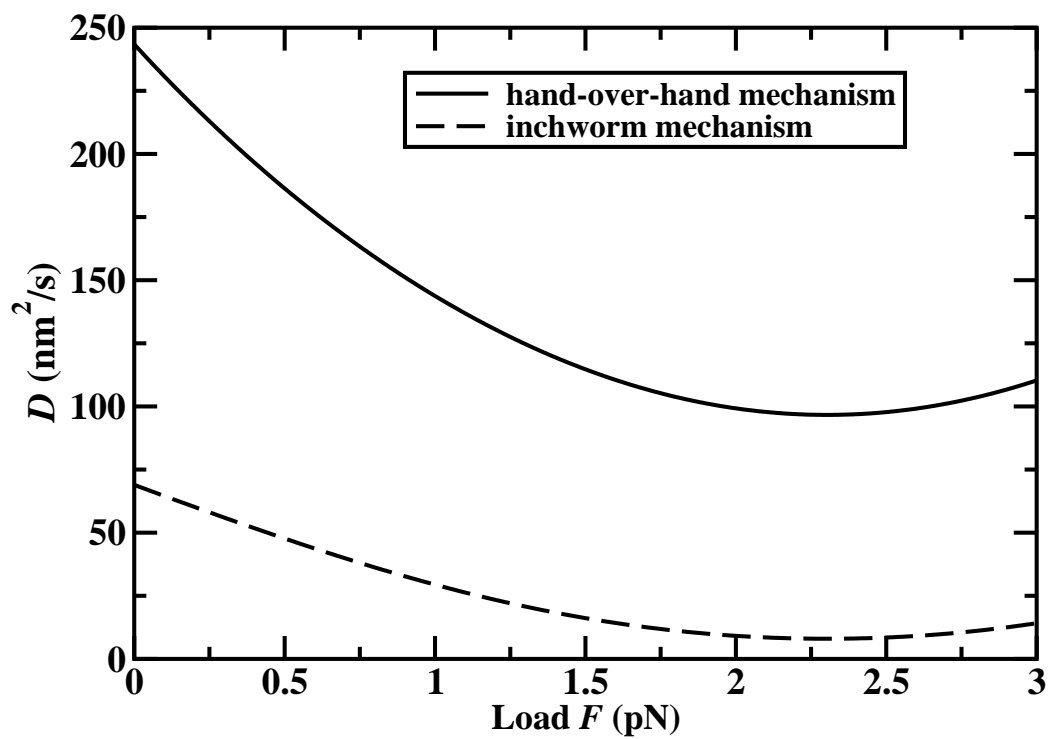}
\caption{Dispersion as a function of external loads for different mechanisms of the motion.  The solid line describes the hand-over-hand mechanism, while the dashed line corresponds to the inchworm mechanism. The parameters are the same as in Fig. 4.}
\label{fig5}
\end{figure}

It is also interesting to compare the dimensionless function randomness for each mechanism: see Fig. 6. These results suggest that the motor proteins in the inchworm mechanism move slower and fluctuates less than the particles in the hand-over-hand mechanism, but the relative decrease in the fluctuations is larger than the relative lowering of the average speed. This observation, that is correct for {\it any} $N$, is very important and it can be used for the experimental discrimination between different mechanisms of motor protein's motility.

We propose the following procedure to determine the mechanism of the motion of motor proteins  using {\it only} experimental measurements. First, from the independent bulk biochemical kinetic experiments determine the number of rate-limiting intermediate states. This information provides the size of the period, i.e., the parameter  $N$. Second, from the high-precision single-molecule trajectories  extract the velocity and dispersion  for different [ATP] and different  external forces. Such data can be obtained from the single-molecule optical trap experiments \cite{schnitzer,kojima,visscher,mehta,schnitzer1,nishiyama}. In the final step, analyze the randomness. If for some  system this procedure yields $r < 1/N$, and the known number of intermediate states is $N$, it indicates that the motor  protein cannot move by the hand-over-hand mechanism. However, for $r > 1/N$ both mechanisms are still possible.

\begin{figure}[tbp]
\centering
\includegraphics[clip=true]{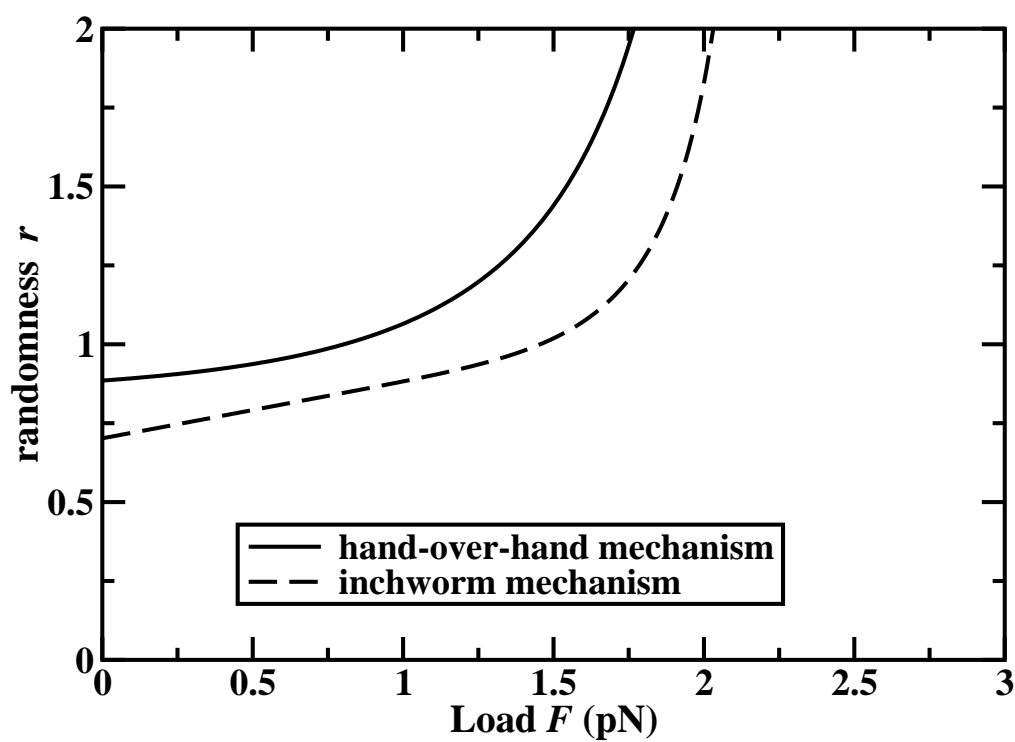}
\caption{Randomness at different external forces.  The solid line describes the hand-over-hand mechanism, while the dashed line corresponds to the inchworm mechanism. The parameters are the same as in Fig. 4.}
\label{fig6}
\end{figure}

\section{Summary and Conclusions}

The dynamics of motor proteins that move along the linear molecular tracks is discussed by taking into account the molecular structure and analyzing in detail two possible mechanisms of motility. The motor proteins are viewed as two interacting particles that correspond to different motor  domains in many conventional molecular motors \cite{howard}. The explicit expressions for the velocity and dispersion are obtained for the hand-over-hand mechanism, when the motor heads pass each other in the alternate fashion, and for the inchworm mechanism, when one motor domain is always ahead of another one.    

The exact calculation of the dynamic properties of molecular motors in the hand-over-hand mechanism are performed by mapping the dynamics of two particles into the one-particle system, for which the dynamic properties are known exactly. It shows that in this case the dynamics is identical to the motion of the single free motor domain on the same biochemical pathway.  The situation is very different for the inchworm mechanism. In this case we derived the exact analytic expressions for the velocity and dispersion by generalizing the single-particle Derrida's method \cite{derrida83,FK,KF00a,KF00b} to the system with two interacting particles. 

Comparing the dynamics of molecular motors in two different modes, we conclude that the proteins in the inchworm mechanism move slower and fluctuate less than the particles in the hand-over-hand mechanism. Our results also indicate that the relative decrease in dispersion, expressed via the randomness parameter, is smaller for the inchworm mechanism. One suggestion is to use this observation for the analysis of experimental data on motor proteins. The method of possible discrimination between two mechanisms of motor protein's  motility based on experimental observations is presented. In addition, the effect of external forces on the dynamic properties of molecular motors in two mechanisms is also discussed. 

The dynamic properties of motor proteins that move through the inchworm mechanism are obtained via the two-particle calculations. However, the average velocity and dispersion could also be obtained by mapping the system with two motor domains into the system with only one particle, for example, the center of mass of the molecule. In general, the inchworm model where the distance between the individual motor domains is not larger than $m$ sites, can be mapped into the motion of a single particle on $m-1$ parallel biochemical pathways, for which the  dynamic properties are known exactly \cite{kolom}.     

Our analysis of motor protein dynamics is rather very simplified since  we  considered the molecules in which subunits interact only  through the hard-core exclusion potential. However, the  heads in motor proteins  coordinate their motion and thus interact much stronger then otherwise might be expected \cite{howard}. It will be interesting to investigate the motor proteins with more realistic interactions between the subunits. The theoretical method used here seems capable of investigating the more realistic systems of  molecular motors.

\ack

 The support from the Camille and Henry Dreyfus New Faculty Awards Program (under Grant No. NF-00-056), from the Welch Foundation (under Grant No. C-1559), and from the US National Science Foundation through the grant CHE-0237105 is gratefully acknowledged.

\end{document}